\documentclass[useAMS,usenatbib]{mn2e}
\usepackage{chngpage}
\usepackage{graphicx}
\usepackage{natbib}
\usepackage{url}
\usepackage{amsmath}
\usepackage{booktabs}
\usepackage{threeparttable}

\usepackage{cite}
\usepackage[usenames,dvipsnames]{color}
\usepackage{amsmath}
\usepackage{amssymb}

\title[Search for a  pulsar in SNR~1987A]
  {Search for a Radio Pulsar in the Remnant of Supernova 1987A}
  \author[Zhang et al.]
    {S.-B. Zhang$^{1,2,3,4}$, S. Dai$^{3}$, G. Hobbs$^{3}$, L. Staveley-Smith$^{4}$, R. N. Manchester$^{3}$, 
\newauthor C. J. Russell$^{5}$, G. Zanardo$^{4}$ and X.-F. Wu$^{1,6}$\\
  $^{1}$Purple Mountain Observatory, Chinese Academy of Sciences, Nanjing 210008, China\\
  $^{2}$University of Chinese Academy of Sciences, Beijing 100049, China\\
  $^{3}$CSIRO Astronomy and Space Science, Australia Telescope National Facility, Box 76, Epping, NSW 1710, Australia\\
  $^{4}$International Centre for Radio Astronomy Research, University of Western Australia, Crawley, WA 6009, Australia\\
  $^{5}$CSIRO Scientific Computing Services, Australian Technology Park, Locked Bag 9013, Alexandria, NSW 1435, Australia\\
  $^{6}$School of Astronomy and Space Science, University of Science and Technology of China, Hefei, Anhui 230026, China\\}

\begin{document}

\maketitle

\begin{abstract}
We have observed the remnant of supernova SN~1987A (SNR~1987A), located in the Large Magellanic Cloud (LMC), to search for periodic and/or transient radio emission with the Parkes 64\,m-diameter radio telescope. We found no evidence of a radio pulsar in our periodicity search and derived 
8$\sigma$ upper bounds on the flux density of any such source of $31\,\mu$Jy 
at 1.4~GHz and $21\,\mu$Jy at 3~GHz. 
Four candidate transient events were detected with greater than $7\sigma$ significance, with dispersion measures (DMs) in the range 150 to  840\,cm$^{-3}\,$pc. For two of them, we found a second pulse at slightly lower significance. However, we cannot at present conclude that any of these are associated with a pulsar in SNR~1987A. As a check on the system, we also observed PSR~B0540$-$69, a young pulsar which also lies in the LMC. We found eight giant pulses at the DM of this pulsar. We discuss the implications of these results for models of the supernova remnant, neutron star formation and pulsar evolution.
\end{abstract}


\section{Introduction}
\label{sec:intro}
SN~1987A, first observed on 1987 February 23, was visually the brightest supernova (SN) since Kepler's SN of 1604 and, being located in the Large Magellanic Cloud (LMC), was the nearest SN in the modern astronomical era. It was also unique in that the progenitor star, a blue supergiant, had been spectroscopically studied before the explosion (see, e.g., Podsiadlowski 1992). It therefore provides a unique opportunity not only to understand Type II supernova explosions, but potentially to also observe an extremely young pulsar. The detection of neutrinos before the optical brightening of SN 1987A (Bionta et al. 1987; Hirata et al. 1987; Alexeyev et al. 1988) supports the theory that a neutron star formed in the core-collapse SN explosion (Arnett et al. 1989). Efforts to search for a pulsar in the supernova remnant (SNR) of SN 1987A (SNR~1987A) have continued since the SN was first detected (Manchester 1988; Kristian et al. 1989).

Initially, optical searches were carried out because the surrounding nebula was expected to be so dense that radio signals would be absorbed, and X-ray telescopes had insufficient sensitivity. Unfortunately, no convincing optical periodicity has been found (Kristian et al. 1989; Murdin et al. 1990; Middleditch et al. 2000) and upper limits (Manchester 1988; Pennypacker et al. 1989; \"Ogelman et al. 1990; Percival et al. 1995, Manchester \& Peterson 1996) were well above the luminosities of similar central compact objects found in other SNR (see Table 4 in Graves et al. 2005 for more details). Recently, analysis of Chandra observations put limits on the X-ray luminosity of a compact object with a non-thermal, Crab-pulsar-like spectrum of the order of $10^{35}$erg s$^{-1}$ (Esposito et al. 2018).

Although scattering, absorption and dispersion in the surrounding nebula might make the radio pulsations hard to detect, a few attempts have been made using the Parkes 64-m radio telescope. Early searches were described by Manchester (1988), who gave an upper limit of about 0.2\,mJy at 1.5\,GHz for a periodic pulse train with pulse periods longer than 10\,ms. Observations have been made every few years since then giving similar upper limits. A more extensive search was carried out in 2006 December (Manchester, 2007). Observational details and upper limits obtained from that study are listed in Table 1 of Manchester (2007). No significant detection was made. The most constraining upper limit of $\sim$50 $\mu$Jy was obtained at 1.5\,GHz assuming a pulse duty cycle of 0.1.

The individual pulses from a given pulsar can vary widely in intensity. Even though most pulsars have been discovered by searching for the periodicity of the pulse train, some pulsars have been discovered through the detection of one, or a few, bright pulses.  The rapidly rotating transient sources (RRATS; McLaughlin et al. 2006) are an extreme example of such pulsars.  Pulsars that exhibit bright single pulses are generally young. The most well-known giant pulse pulsar, the Crab pulsar, can generate pulses whose peak flux densities are more than million times brighter than its normal radio pulses (Hankins \& Eilek 2007). It is therefore possible that the pulsar in SN 1987A also emits individual bright pulses, which may enable us to identify the pulsar.

In 2013, we carried out an extensive set of observations with the Parkes telescope with the primary goal of again searching for the elusive pulsar. We also observed the pulsar PSR~B0540$-$69. This well-studied pulsar is within SNR~0540$-$693 (Manchester, Staveley-Smith \& Kesteven 1993a) and is also located in the LMC, not far from SN 1987A (with an angular separation of $\sim 1.2^\circ$). PSR~B0540$-$69 has been detected at X-ray (Seward, Harnden, \& Helfand 1984), optical (Middleditch \& Pennypacker 1985) and radio wavelengths (with a weak and broad radio pulse at 0.64 GHz discovered by Manchester et al. 1993b). Giant pulses have also been detected from this pulsar at 1.4\,GHz, with flux densities up to 5000 times the mean single-pulse value (Johnston \& Romani 2003).  

In this paper, we present the results of our search for periodic and single-pulse signals from both SNR~1987A and PSR~B0540$-$69.  Details of the observations and data reduction are given in the following section. In Section~\ref{sec:result}, we present the results of single-pulse and periodicity searches, and highlight the most significant candidates in our search for individual pulses. A discussion of our results is given in Section~\ref{sec:dis} and we conclude in Section~\ref{sec:conc}.         

\section{Observation and Data Reduction}
\label{sec:data}
\subsection{Observations}
\label{sec:obs}
The observations used in this paper were obtained with the Parkes 64\,m-diameter radio telescope on 2013 May 8, 24 and 25, and on 2013 October 6--7. 
We used the 10/50\,cm dual-band receiver providing simultaneous data streams centred at $732$ and $3100$\,MHz on May 8 and October 6. On the other days (May 24, 25 and October 7) we observed using the central beam of the 20\,cm multibeam receiver centred at $1369$ MHz. PSR~B0540$-$69 was observed for 3.23 hours using the central beam of the multibeam receiver. The channelised signals were two-bit sampled and recorded using Parkes Digital Filterbank (PDFB) systems. The bandwidth, number of channels, sampling time and integration time of these observations are listed in Table~\ref{table:parameters}. More details of the observing and data recording systems are available in Manchester et al. (2013). 

All data obtained for this project are available for download from CSIRO's\footnote{Commonwealth Scientific and Industry Research Organisation} Data Access Portal (DAP)\footnote{\url{https://data.csiro.au}; see Hobbs et al. (2011) for details.}. The data were recorded under the P834 observing code and are available for download\footnote{e.g., Staveley-Smith, Manchester \& Zanardo (2014a) and Staveley-Smith, Manchester \& Zanardo (2014b). Note that some observations in the latter collection are still under embargo. The data files will become public 18 months after the date of the observation.}. The PDFB4 backend instrument at Parkes splits each search-mode observation into multiple data files in order to ensure that no individual file becomes too large. The data files corresponding to each observation were therefore concatenated using the \emph{\sc addsearch} routine prior to carrying out periodicity searches.

\begin{table*}
\caption{Observation summary and results of a periodicity search for the possible pulsar of SNR~1987A}
\begin{center}
\begin{tabular}{ccccccccccl}
\hline
\hline
Source & Freq. & Date & Total/Chan. & $t_{\rm samp}$ & Integration & $T_{sys}$ & G & ${\rm DM}\, {\rm Range}$ &  8$\sigma$ Limit    \\
       &   (MHz) &  (MJD)  & BW(MHz)     & ($\mu$s)      & Time(h)  & (K) & (K/Jy) & $($cm$^{-3}\,$pc$)$          & $(\mu {\rm Jy})$ \\
\hline
SNR~1987A & 1369 & 56437 & 256/0.5 & 100 & 6.50 & 21 & 0.735 & 0-2000 & 33\\
SNR~1987A & 1369 & 56438 & 256/0.5 & 96  & 7.27 & 21 & 0.735 & 0-2000 & 31\\
SNR~1987A & 1369 & 56573 & 256/0.5 & 96  & 7.17 & 21 & 0.735 & 0-2000 & 31\\
SNR~1987A & 732  & 56420 &64/0.125 & 96  & 1.50 & 40 & 0.909 & 0-2000 & 212\\
SNR~1987A & 3100 & 56420 &1024/2.0 & 96  & 1.50 & 35 & 0.909 & 0-2000 & 46\\
SNR~1987A & 732  & 56572 &64/0.125 & 96  & 7.17 & 40 & 0.909 & 0-2000 & 97\\
SNR~1987A & 3100 & 56572 &1024/2.0 & 96  & 7.17 & 35 & 0.909 & 0-2000 & 21\\
PSR~B0540$-$69  & 1369 & 56437$-$56572 & 256/0.5  & 100/96 & 3.23 & 21 & 0.735 & 0-200  & - \\
\hline
\end{tabular}
\end{center}
\label{table:parameters}
\end{table*}

\subsection{Single Pulse Search}
\label{sec:sing}
We used the pulsar searching software package \emph{\sc presto}\footnote{\url{http://www.cv.nrao.edu/~sransom/presto/}} (Ransom 2001). Since the data have been recorded applying {\sc psrfits} scales and offsets, we used the options \emph{\sc -noscales} and \emph{\sc -nooffsets} during the processing. We processed the data on the CSIRO's high performance computer (HPC) facilities. Strong narrow-band and short-duration broadband radio frequency interference (RFI) were identified and marked using the \emph{\sc presto} routine \emph{\sc rfifind} and a mask file was produced for each observation. We used a 1\,s integration time for our RFI identification and a $6 \sigma$ cutoff to reject time-domain and frequency-domain interference in our pipeline. In preparation for de-dispersion, the \emph{\sc ddplan.py} algorithm was used to determine the dispersion measures (DMs) required for us to search.
The DM range that we searched was $0-2000\,\,$cm$^{-3}\,$pc and the numbers of the DM trials for data at $732$, $1369$ and $3100$\,MHz were 2,773, 2,244 and 1,872,
respectively. We did not search for higher DMs because the pulse is significantly smeared within channels and the sensitivity is reduced, especially for short-period pulsars. Data were then de-dispersed at each of the trial DMs using the \emph{\sc prepdata} routine, and RFI was removed based on the mask file. Single pulse candidates with a signal-to-noise ratio (S/N) larger than five were identified using the \emph{\sc single\_pulse\_search.py} routine for each de-dispersed time series file, and boxcar filtering with width up to 300 samples was used (filter widths of 1, 2, 3, 4, 6, 9, 14, 20, 30, 45, 70, 100, 150, 220 and 300 samples).

Our pipeline generated 10,355,344 candidates with a S/N $\ge 5$ and 1,408,000 candidates with a S/N $\ge 7$ for the whole SNR~1987A data set. To reduce the number of candidates to inspect, we grouped all the candidates that occurred close together in time (within a 10\,ms time-window) and then only inspected the highest S/N one. This led to a reduced list of 10,990 candidates, which were then visually inspected.

\subsection{Periodicity Search}
\label{sec:per}
The periodicity search was also carried out using \emph{\sc presto}. 
In a similar manner to the single pulse searches, RFI was rejected and marked using \emph{\sc rfifind} and the trial DMs up to $2000\,\,$cm$^{-3}\,$pc were determined using  \emph{\sc ddplan.py}. 
The concatenated data were then de-dispersed at each trial DM using the \emph{\sc prepsubband} software which also removes Doppler shifts caused by the Earth's motion.   
The resulting time series were Fourier transformed (using the  \emph{\sc realfft} package).
Periodic signals were searched for in the Fourier domain using \emph{\sc accelsearch}.  We did not carry out a search for a highly accelerated system.
Candidates less than 4$\sigma$ (from incoherent power summation) and with maximum S/N occuring at a DM of 0\,cm$^{-3}$pc were rejected. Multiple detections of the same candidates (e.g. the harmonics and the same barycentric period at multiple DMs) were combined using \emph{\sc accel\_sift.py}. 
Finally, candidates with S/N larger than 8 suggested by \emph{\sc accel\_sift.py} were folded using \emph{\sc prepfold}.

\section{Results}
\label{sec:result}

\subsection{Single Pulse Search}
\label{sec:B0540}\label{sec:S1987A}
The 3.23\,hr observation of PSR~B0540$-$69 was processed using the pipeline described in Section~\ref{sec:sing}. As the DM of PSR~B0540$-$69 is known to be about $146.5\,\,$cm$^{-3}\,$pc (Manchester et al. 1993a; Johnston \& Romani 2003), we simplified our pipeline and only searched for DMs up to $200\,\,$cm$^{-3}\,$pc. 

\begin{figure*}
\begin{center}
\includegraphics[scale=0.6]{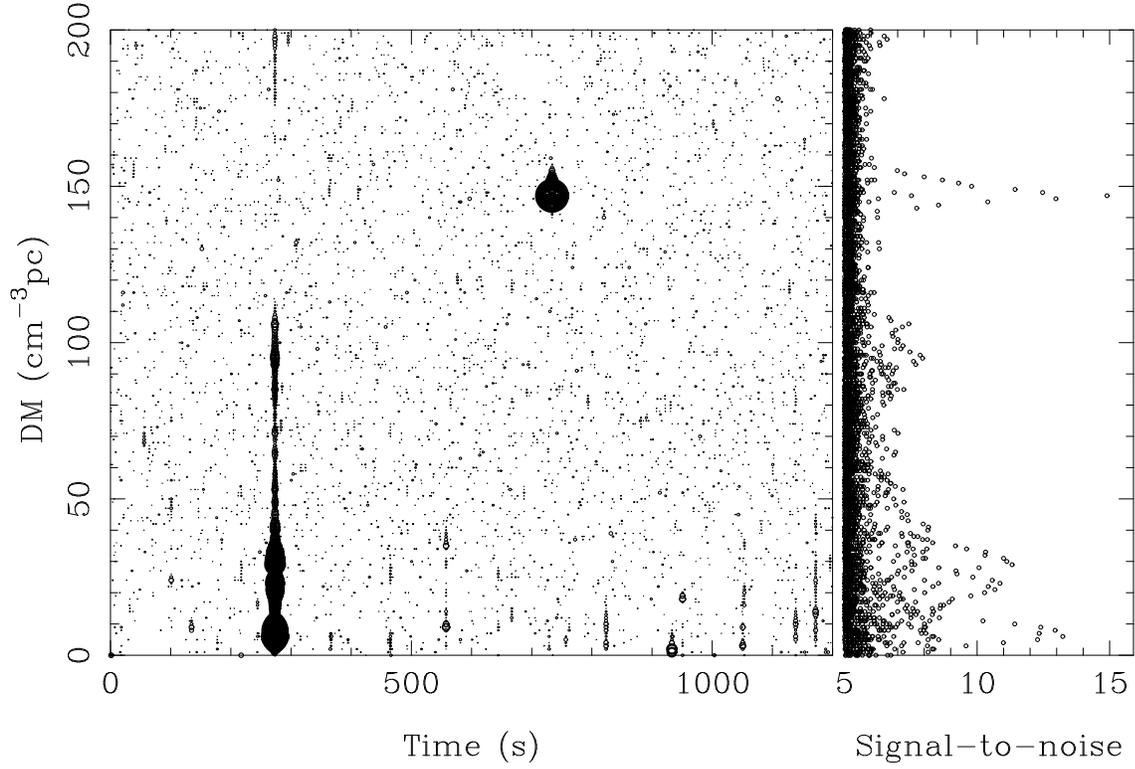}
\caption{Output of our processing pipeline for an observation of PSR~B0540$-$69. Each candidate pulse with a S/N$>$5 is shown as a filled circle in the left-hand panel, with diameter proportional to S/N. One giant pulse was detected at $147$cm$^{-3}\,$pc, corresponding to the known DM of this pulsar. The right-hand panel displays each detected pulse as a function of S/N and DM.} 
\label{sgp:B0540}
\end{center}
\end{figure*}

\begin{figure*}
\begin{center}
\begin{tabular}{ll}
\includegraphics[width=6.5cm,angle=-90]{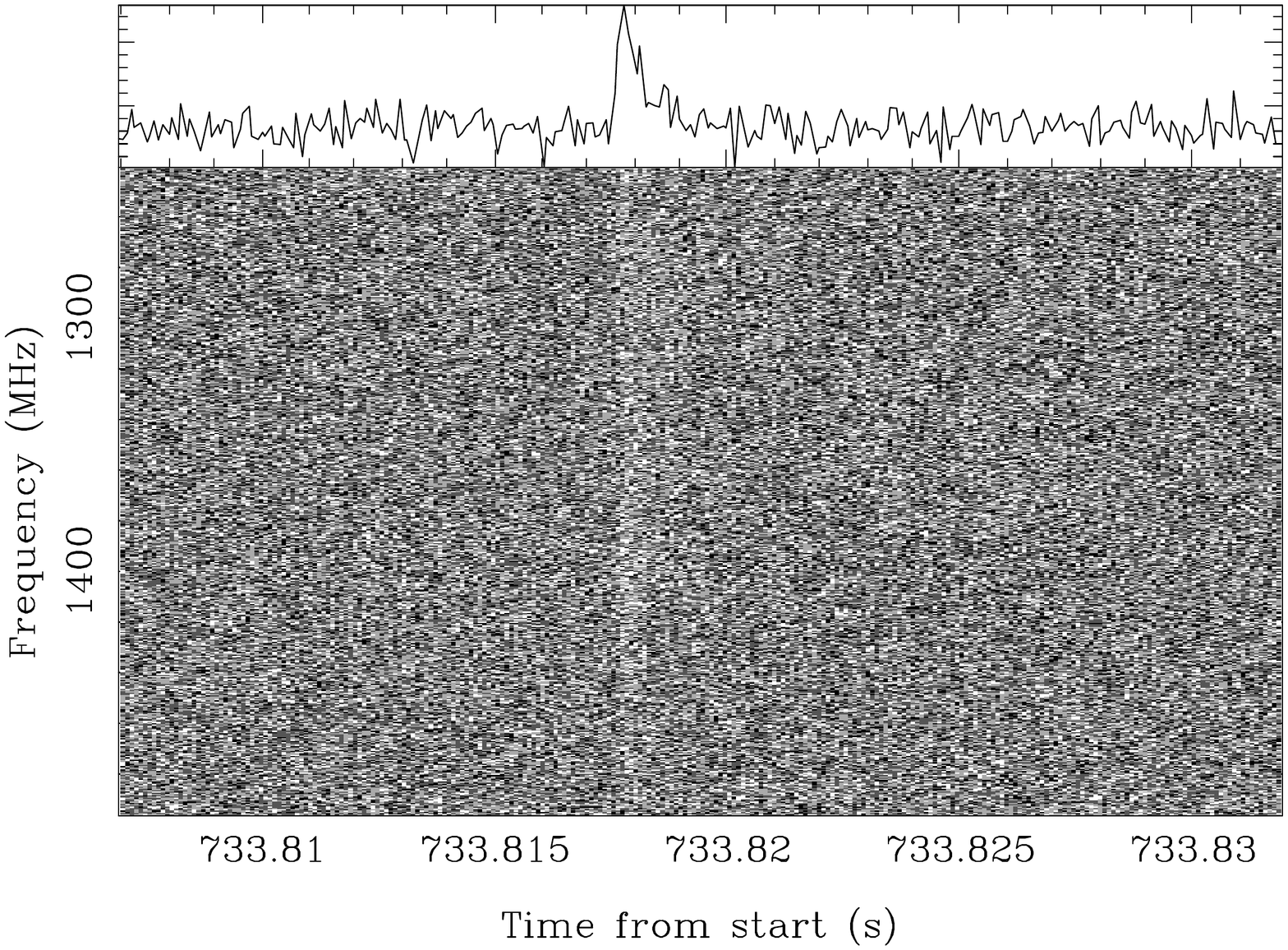} &
\includegraphics[width=6.5cm,angle=-90]{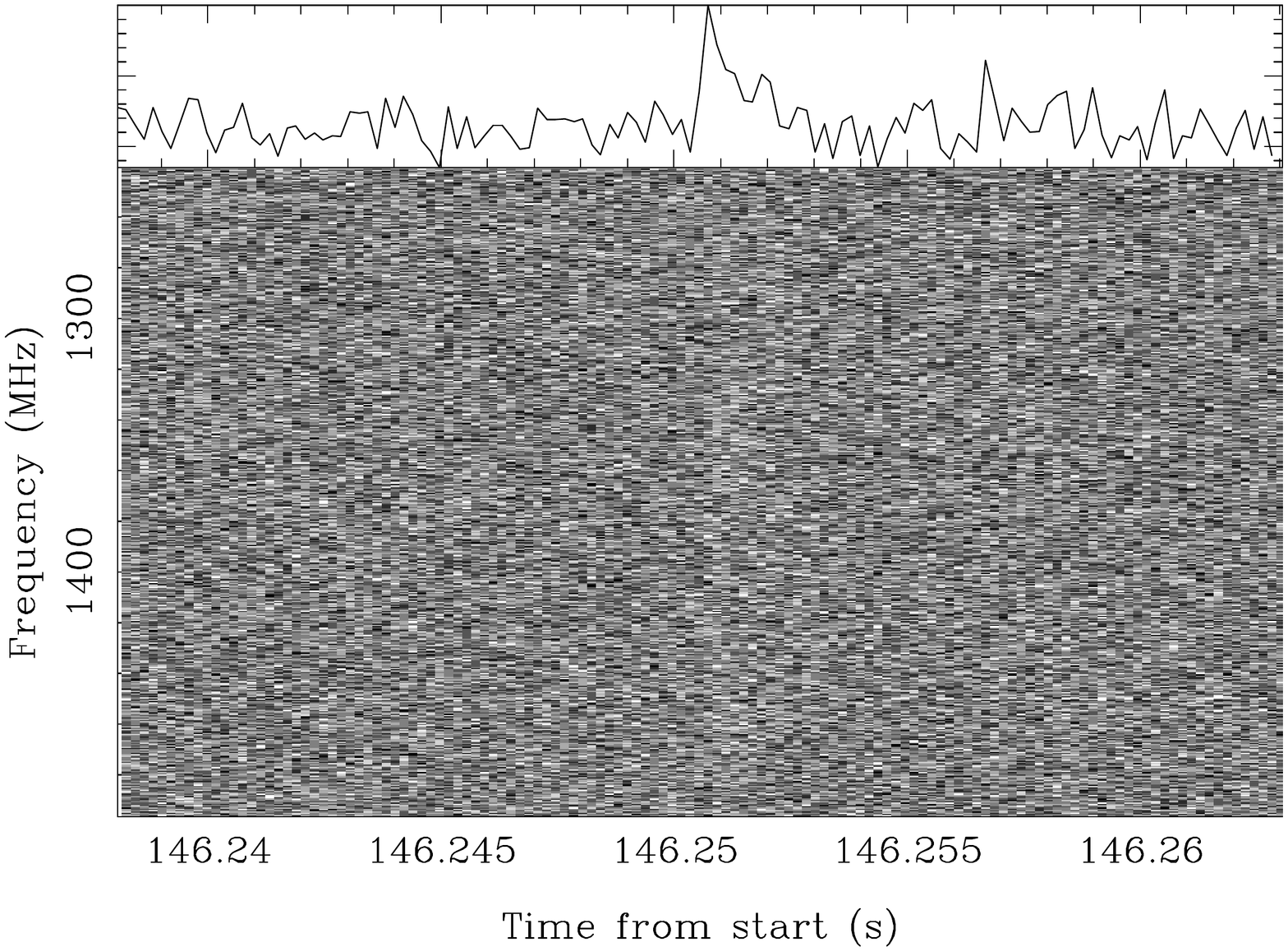} \\
\end{tabular}
\caption{Frequency-time planes and integrated pulse shapes of the brightest (left) and the weakest (right) detected giant pulses from PSR~B0540$-$69. The pulses are de-dispersed at the optimal DM and the y-axis frequency range is 1241 MHz to 1497 MHz on both plots.} 
\label{B0540}
\end{center}
\end{figure*}

At a DM of $147\,\,$cm$^{-3}\,$pc, eight giant pulses were detected in our data, which gives an event rate consistent with that published in Johnston et al. (2004). An example of a giant pulse detected from PSR~B0540$-$69 is shown in Figure~\ref{sgp:B0540}, with the highest S/N occurring at the known DM of PSR~B0540$-$69. The brightest giant pulse has a S/N of 14.6 (left panel of Figure~\ref{B0540}) and the weakest pulse has a S/N of 7.5 (right panel of Figure~\ref{B0540}). These observations allowed us to confirm that our pipeline was working and is capable of detecting pulses from pulsars at the distance of the LMC. 

From the 10,990 candidates, which were inspected by eye for the SN~1987A search, we identified four plausible pulses. 
Table~\ref{table:single:candidate} tabulates the times and DMs of these candidates and they are shown in Figures~\ref{Cand_SNR}. These four candidates all have different DMs (from 151 to 839\,cm$^{-3}\,$pc). As the DM to SN 1987A is unknown, in an attempt to confirm these candidates, we re-analyzed the data to search for candidates with S/N$\ge 6$ in a DM of range $\pm$5\,cm$^{-3}\,$pc centered at the candidate DMs and with a DM step of 0.1\,cm$^{-3}\,$pc.  Two plausibly related pulse candidates were found (candidate 3A and 4A in Table~\ref{table:single:candidate} and are shown in the right panels (middle and bottom) of Figures~\ref{Cand_SNR}). These were discovered in the same observing session as candidate 3 and 4 (2013 October 7). The candidate pulses occur at DMs of 560 and 151\,cm$^{-3}\,$pc, respectively.

\begin{table}
\label{T_cand}
\caption{Single-pulse candidates for SNR~1987A, all obtained in the 20\,cm observing band}
\begin{center}
\begin{threeparttable}
\begin{tabular}{cccccccccccl}
\hline
\hline
Cand. & Obs. Date & S/N & ${\rm DM}$      & boxcar  \\
Number    &  (MJD)   &    & $($cm$^{-3}\,$pc$)$ & (bins) \\
\hline
1 & 56438.07 & 7.0 & 409.0$\pm$0.1 & 2 \\ 
2 & 56438.17 & 7.2 & 839.0$\pm$0.1 & 1 \\ 
3 & 56572.60 & 7.1 & 560.0$\pm$0.6 & 3 \\
3A\tnote{a} & 56572.75 & 6.7 & 559.9$\pm$0.2 & 2 \\  
4 & 56572.68 & 7.0 & 151.0$\pm$0.3 & 6 \\ 
4A\tnote{a}& 56572.83 & 6.6 & 150.8$\pm$0.2 & 3 \\ 
\hline
\end{tabular}
     \begin{tablenotes}
        \footnotesize
        \item[a]Further candidates found with S/N $\ge 6$ at nearby DMs.
     \end{tablenotes}
\end{threeparttable}
\label{table:single:candidate}
\end{center}
\end{table}

\subsection{SNR~1987A Periodicity Search}
\label{sec:P1987A}
No convincing candidate with a S/N greater than 8 was detected using our periodicity-search pipeline \footnote{Note that Manchester (2007) used a 9$\sigma$ threshold when selecting pulsar candidates}, and no significant periodic signal was found at the DM of any of our single pulse candidates. Limits on the flux density can be estimated as:
\begin{equation}
S_{lim}=\frac{\sigma\,{\rm S/N}_{\rm min}\,T_{\rm sys}}{G \sqrt{{\Delta}{\nu}N_p{t}_{\rm obs}}}\sqrt{\frac{\delta}{1-\delta}}, 
\label{equ:limit}
\end{equation}
where $\sigma = 1.5$ is a loss factor (Manchester et al. 2001), $T_{\rm sys}$ is the system temperature and $G$ is the telescope antenna gain\footnote{$T_{\rm sys}$ and $G$ for the 10/50\,cm receiver and the multibeam receiver can be found from                            \url{https://www.parkes.atnf.csiro.au/observing/documentation/user_guide/pks_ug_3.html#Sensitivity}}, ${\Delta}{\nu}$ is the observing bandwidth, $N_p$ is the number of polarizations, $t_{obs}$ is the integration time and $\delta$ is the pulse duty cycle (assumed to be 0.1). We used ${\rm S/N}_{\rm min} = 8.0$ as a threshold for the cut-off signal-to-noise ratio. The flux density limits at different frequencies and for different observations are presented in the last column of Table~\ref{table:parameters}.

\section{Discussion}
\label{sec:dis}

\begin{figure*}
\begin{center}
\begin{tabular}{ll}
\includegraphics[width=6.5cm,angle=-90]{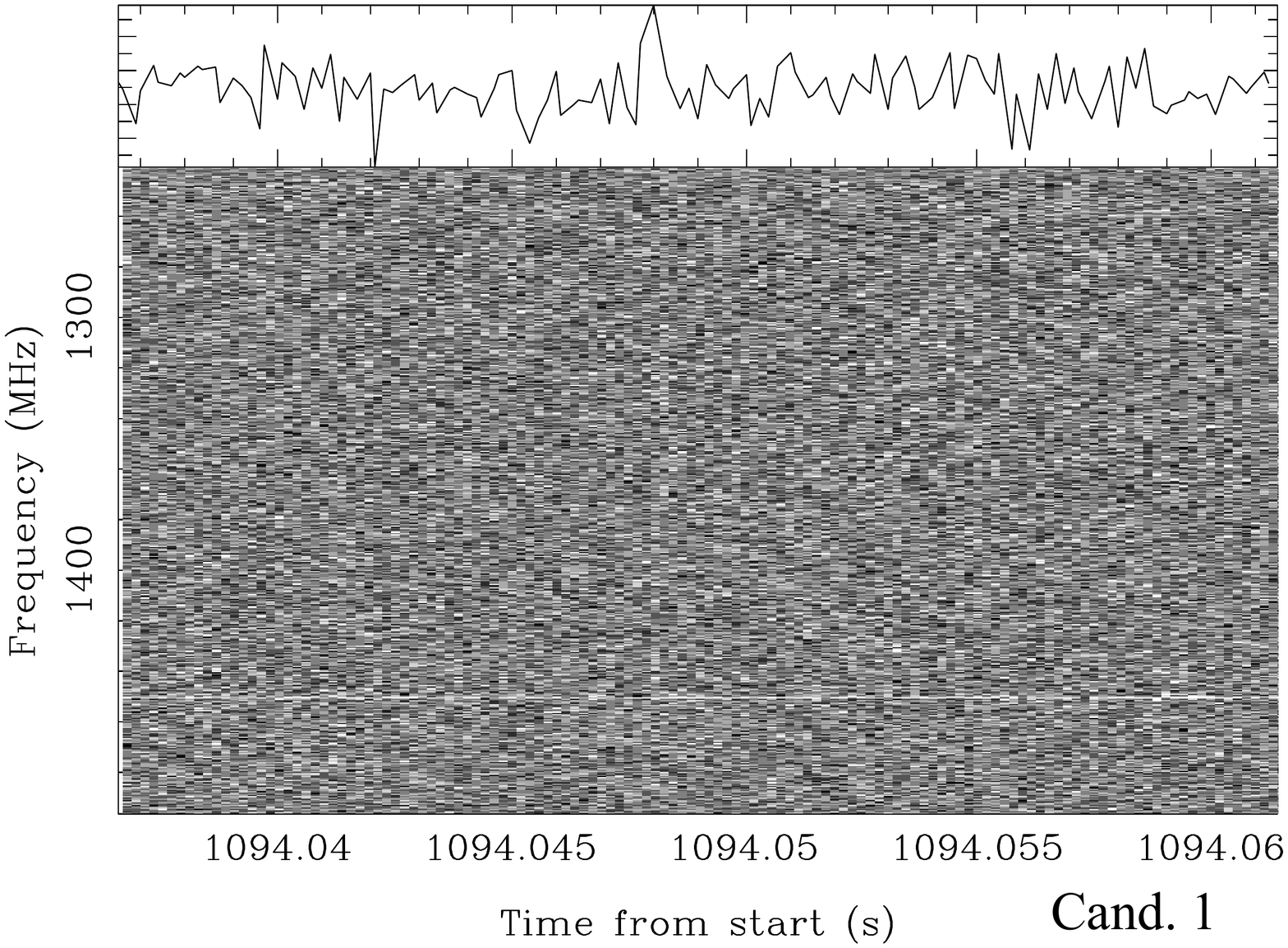} & \includegraphics[width=6.5cm,angle=-90]{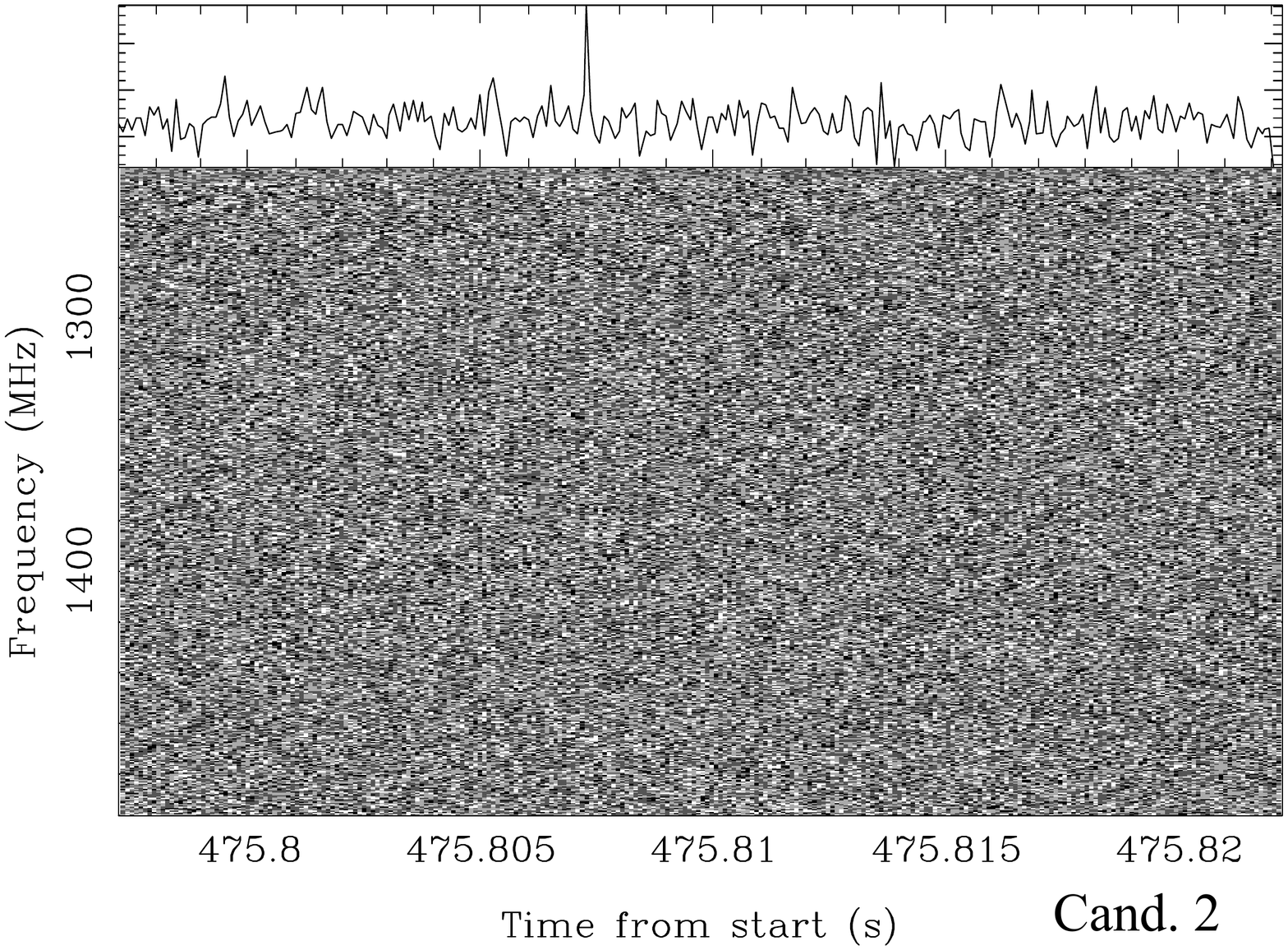} \\
\includegraphics[width=6.5cm,angle=-90]{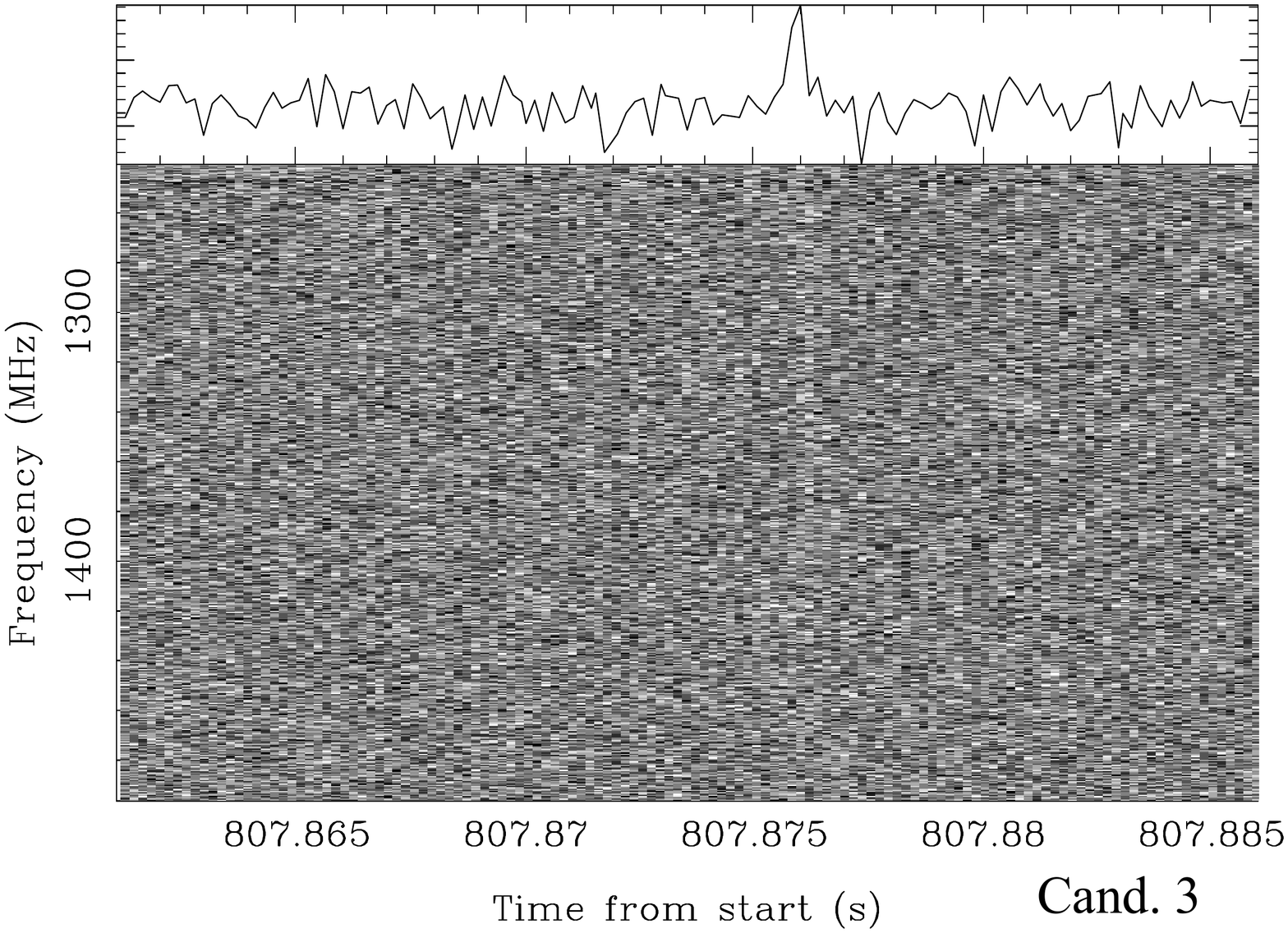} & \includegraphics[width=6.5cm,angle=-90]{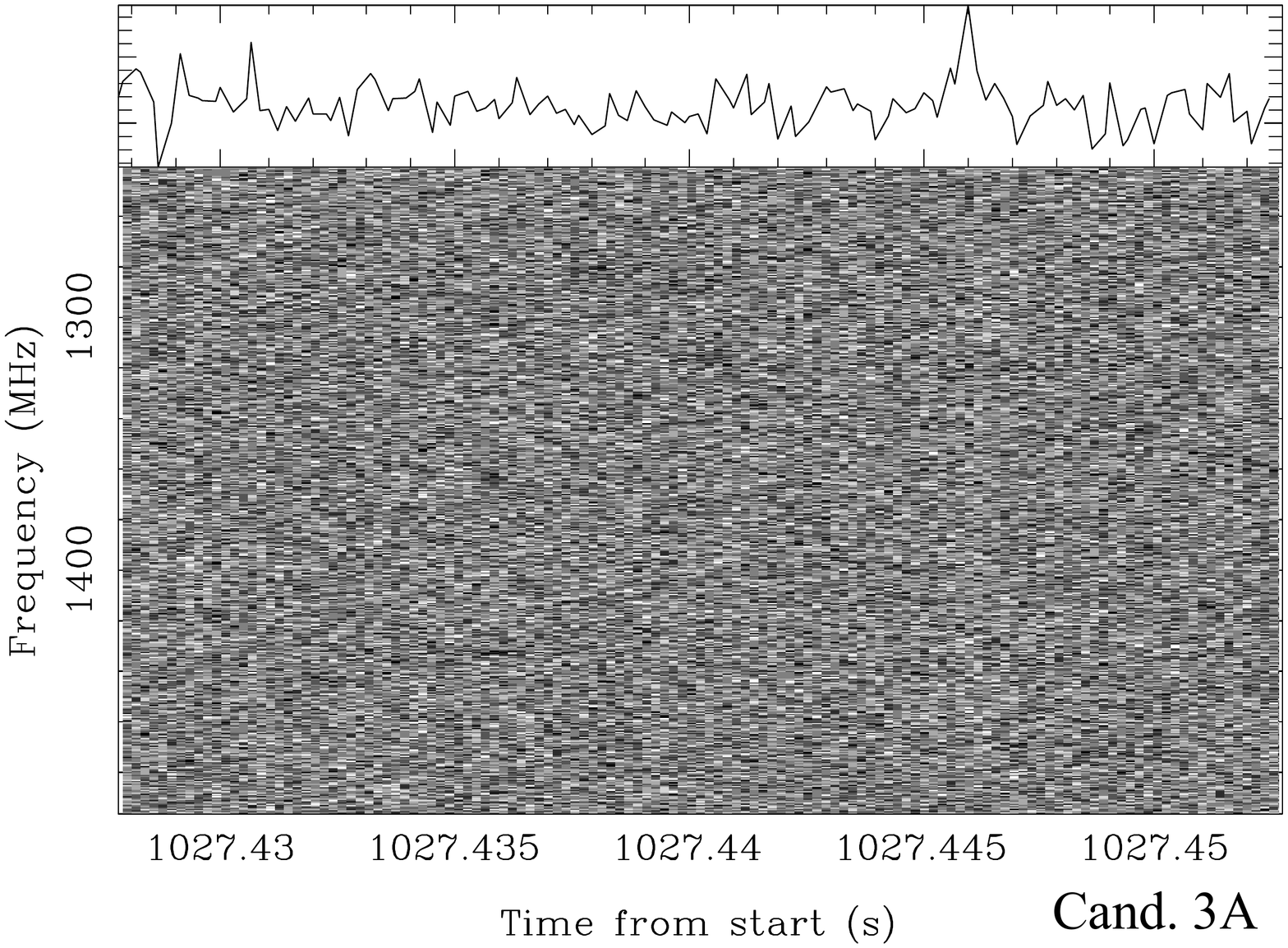} \\
\includegraphics[width=6.5cm,angle=-90]{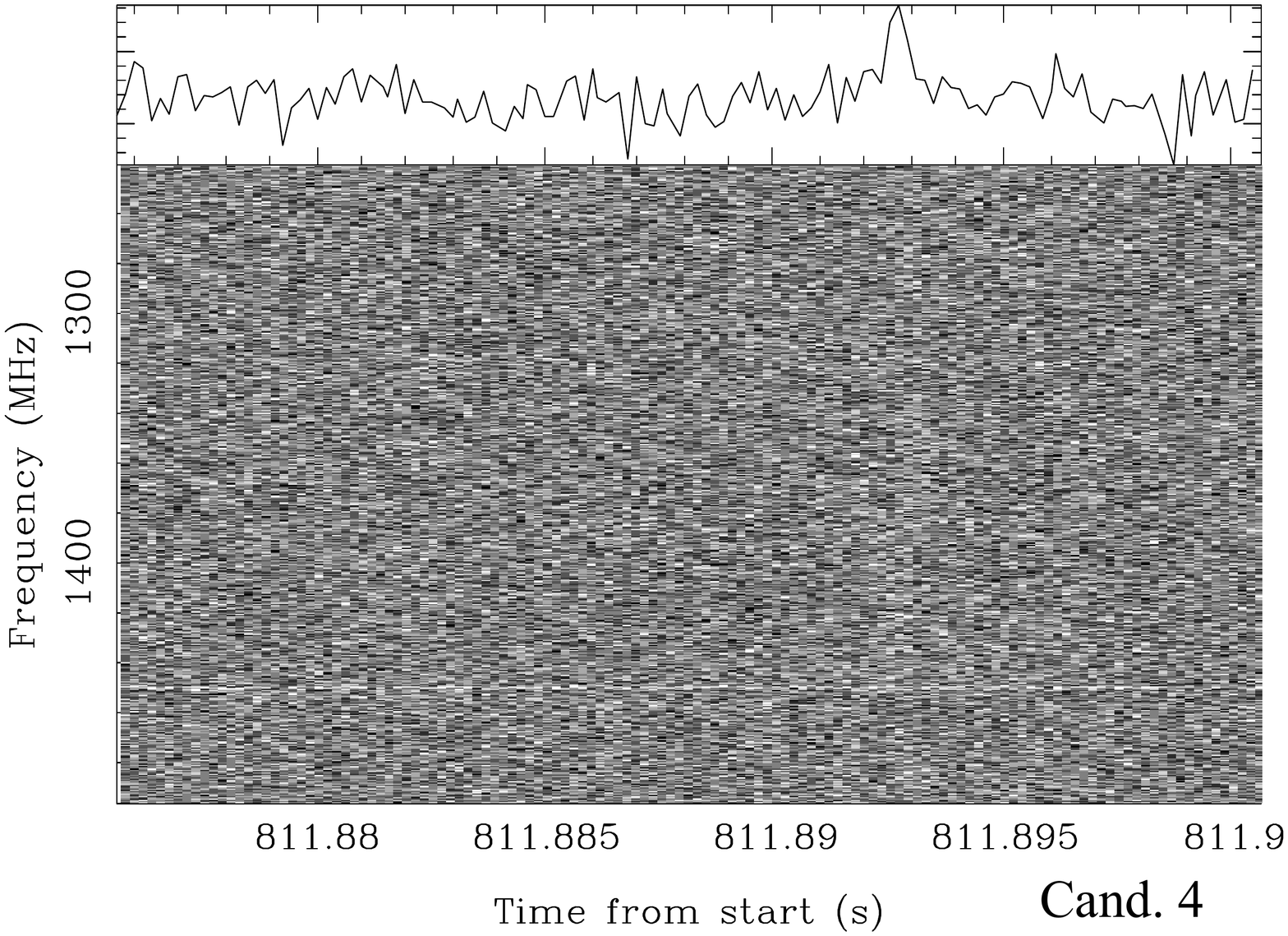} & \includegraphics[width=6.5cm,angle=-90]{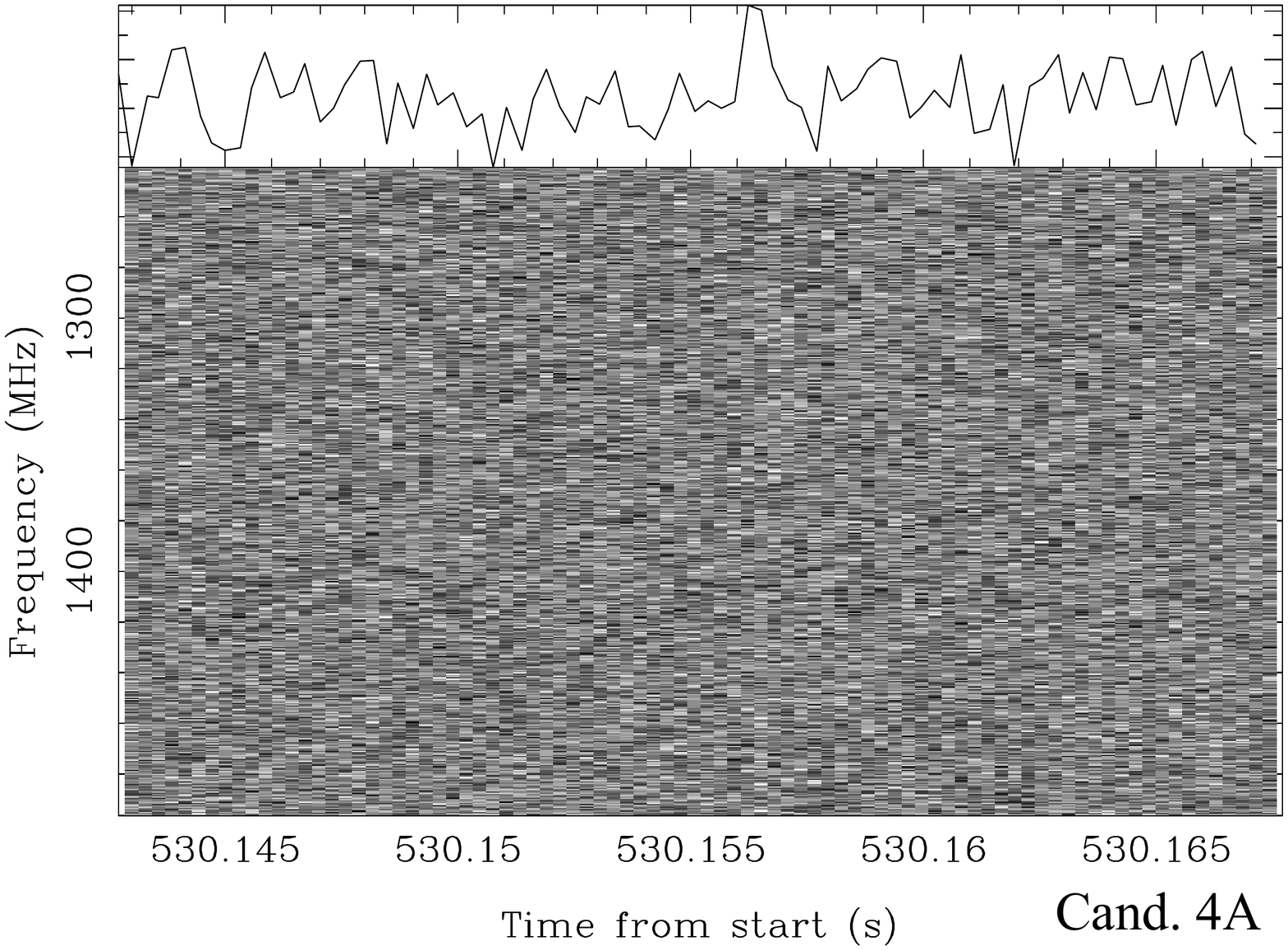} \\
\end{tabular}
\caption{SNR~1987A single pulse candidates listed in Table~\ref{table:single:candidate}. The pulses are de-dispersed at the optimal DM and the y-axis frequency range is 1241 MHz to 1497 MHz on all the plots.} 
\label{Cand_SNR}
\end{center}
\end{figure*}

The candidate pulses from the single pulse search all have S/N around 7 and DMs from 150 to 840\,cm$^{-3}$pc. For comparison the YMW16 electron density model (Yao, Manchester \& Wang 2017) predicts that, in the direction of SN 1987A, the contribution to the DM from our Galaxy and the LMC are about 61 and 218\,cm$^{-3}\,$pc, respectively. PSR~B0540$-$69 has a DM of about 146.5 $\,\,$cm$^{-3}\,$pc (Johnston \& Romani 2003), somewhat smaller than the model prediction. However, the DM measured from any pulsar within SNR~1987A may be significantly larger because of the dense medium surrounding the pulsar. Zanardo (2014a) estimated that the most likely DM associated with a possible pulsar in SNR~1987A is from 100$\,\,$cm$^{-3}\,$pc to 6000$\,\,$cm$^{-3}\,$pc, assuming the ionised hydrogen mass of the ejecta to be $0.1M_{\odot} \leqslant M_{\rm HII} \leqslant 2.5M_{\odot}$ and uniformly distributed within a spherical region of radius $0.01 \leqslant R_{\rm HII} \leqslant 0.10$ pc. Any of the measured DMs are plausible for a pulsar within SN~1987A, but it is clear that they cannot all be from such a pulsar. Currently we have no way of telling which if any of the detected pulses are from a pulsar in SNR~1987A.

Considering the low S/N of our candidates, the fact that candidates were found at different DMs, and that we saw repeated pulses at two very different DMs, it is not possible for us to claim any detection of a pulsar in SNR~1987A.  The candidates may well come from statistical noise fluctuations\footnote{To explore this possibility we simulated short, 120\,s, data sets, but with only white noise. We ran our pipeline and inspected the resulting candidates.  Candidates with S/N$\sim$6 were identified and so it is likely that our actual candidates (which are from data sets much longer than the simulations) are statistical fluctuations.} or low-level RFI. However, we note that candidate 3 (Figure~\ref{Cand_SNR}) can clearly be seen in the time-frequency image and is more obvious than the weakest giant pulse found in PSR~B0540$-$69. Further observations are necessary to confirm its nature.

A number of theories have  been proposed to explain why no detection of a pulsar has yet been made.  Bethe \& Brown (1995) and Chevalier (1996) argued that the neutron star could have accreted matter and collapsed to a black hole.  However, Fryer (1999) showed that such an accretion-induced collapse is very unlikely as the maximum mass of the SN~1987A compact remnant, which is estimated from the mass of ejected $^{56}{\rm Ni}$, is insufficient to collapse into a black hole. 

It has also been suggested that the pulsar magnetic field may take decades to develop (Blandford \& Romani 1988; Bonanno et al. 2005), so the pulsar could still be undetectable. Another possibility is that the pulsar is simply not luminous enough to be detected. The median 1.4~GHz radio luminosity $L_{\rm R,1.4} = S_{1.4} d^2$, where $S_{1.4}$ is the mean 1.4 GHz pulsed flux density and $d$ is the pulsar distance, for known Galactic pulsars is about 9~mJy~kpc$^2$.\footnote{ATNF Pulsar Catalogue, V1.58, Manchester et al., 2005} In contrast, the mean $L_{\rm R,1.4}$ for Magellanic Cloud radio-detected pulsars is about 300~mJy~kpc$^2$. Radio luminosity is not a strong function of pulsar age -- for pulsars with characteristic ages $<10^5$~yrs, the median luminosity is about 12~mJy~kpc$^2$ -- and so these luminosities are relevant to the expected pulsar in SNR~1987A. 
Our 1.4~GHz limit of 31~$\mu$Jy corresponds to a radio luminosity of about 75~mJy~kpc$^2$, relatively high by the standards of Galactic pulsars. However, we note that PSR~B0540$-$69 has a smaller flux density and radio luminosity, about 24~$\mu$Jy (Johnston et al. 2004) and 58~mJy~kpc$^2$, respectively.

Even if the putative pulsar is bright enough, the radiation may not be beamed toward us. The beaming fraction of a pulsar, defined as the fraction of the sky covered by the radiation beam, is typically assumed to be $\sim$20\% in the radio band (Manchester 2007). For young pulsars, Frail \& Moffett (1993) argued the beaming fraction can be as large as $60\%$ based on the sensitive and high-resolution images taken by the Very Large Array (and see similar results in Ravi, Manchester \& Hobbs 2010). However, observations show that the radio beam is often ``patchy" (Lyne \& Manchester 1988; Manchester 1995; Han \& Manchester 2001), and therefore our line of sight to the pulsar may miss strong parts of the beam.

Zanardo (2014a) also showed that it is probable that the surrounding nebula remains too dense for radio signals to escape.  However, the detectability rapidly increases with time. It is therefore essential that we keep searching for the pulsar signal. However, noting our limited knowledge of the density distribution of the inner ejecta and the asymmetry and clumpiness of the medium (Reynolds et al. 1995; Kj{\ae}r et. 2010; Zanardo et al. 2014b; Abell{\'a}n et al. 2017), it is difficult to estimate the earliest time that the pulsar may be expected to appear at radio wavelengths. The anomalous abundances in the equatorial ring (ER) and ejecta asymmetries (e.g. Larsson et al. 2013) could be due to another SN explosion that occurred about 100,000 yr before 1987 in the binary system that was the progenitor of SN 1987A (De Loore \& Vanbeveren 1992).

\section{Conclusions}
\label{sec:conc}   

We have searched for individual pulses from the remnant of SN~1987A with the Parkes radio telescope in several data sets of up to 7~hr duration at radio frequencies about 0.7~GHz, 1.4~GHz and 3~GHz. Four candidate pulses with ${\rm S/N}\geq 7\sigma$ were detected, with DMs in the range of 150 to 840\,cm$^{-3}$~pc. Detections of further pulses at one of the candidate  DMs will be required to confirm if any of our candidate pulses are from a pulsar in SNR~1987A. The observing and processing systems were verified by the detection of eight giant pulses from the Magellanic Cloud pulsar PSR~B0540$-$69. We also undertook a periodicity search on the data sets, with no signficant detection. The upper limits on the mean pulsed flux density at 1.4~GHz and 3~GHz are 31~$\mu$Jy and 21~$\mu$Jy, respectively. The 1.4~GHz upper limit is higher than the measured mean pulsed flux density of PSR~B0540$-$69. Because of this and the uncertainty about the origin of the detected individual pulses, there is a clear case for more radio searches, preferably with higher sensitivity. 

As the pulsar and supernova remnant evolve, the possibility of detecting the pulsar rapidly increases (note that the Crab pulsar was detected less than 1000\,yr after its birth and was likely to be potentially detectable much earlier). The effect of the surrounding nebula is highly frequency dependent, and highly dependent on viewing geometry.  The forthcoming installation of a new wideband (0.7 to 4 GHz) receiver on Parkes telescope will increase the possibility of detection. The higher sensitivity of the receiver, because of its lower system temperature and larger bandwidth, will enable weaker pulses to be detected. The higher frequencies accessible to the receiver will also permit us to probe deeper into the nebula.

\section*{Acknowledgments}
The Parkes radio telescope is part of the Australia Telescope National Facility which is funded by the Australian Government for operation as a National Facility managed by CSIRO.  This paper includes archived data obtained through the CSIRO Data Access Portal (http://data.csiro.au). This work was supported by a UCAS Joint PhD Training Program grant and the National Natural Science Foundation of China (Grant No. 11725314). We acknowledge discussions and support from Lawrence Toomey and Simon Johnston.

\end{document}